\documentclass[12pt]{article}

\usepackage{graphicx}
\usepackage{subcaption}
\usepackage{amsmath}
\usepackage{dsfont}
\usepackage{placeins}
\usepackage{amssymb}
\usepackage{abstract}
\usepackage[auth-sc,affil-sl]{authblk}
\usepackage{hyperref}
\usepackage{apacite}
\usepackage{enumerate}
\usepackage{fancyhdr}
\usepackage[xcdraw,table]{xcolor}
\usepackage{natbib}
\usepackage{algorithm}
\usepackage{algorithmicx}
\usepackage{algcompatible}
\usepackage{algpseudocode}
\usepackage[table]{xcolor}
\usepackage{multirow}
\usepackage{amsthm}

\usepackage{algpseudocode}
\usepackage{amsthm}
\usepackage{hyperref}
\usepackage{mathrsfs}
\usepackage{amsfonts}
\usepackage{bbding}
\usepackage{listings}
\usepackage{appendix}
\usepackage[margin=0.98in]{geometry}
\usepackage{fancyvrb}
\usepackage{cancel}
\newtheorem{theorem}{Theorem}[section]
\newtheorem{remark}{Remark}[section]

\newcounter{probnum}
\allowdisplaybreaks

\definecolor{tabblue}{rgb}{.870588,.905882,.94902}
\definecolor{gray}{rgb}{0.7,0.7,0.7}
\definecolor{black}{rgb}{0,0,0}
\definecolor{white}{rgb}{1,1,1}
\definecolor{blue}{rgb}{0.0,0.0,1}
\newcommand{\ingray}[1]{\color{gray}#1\color{black}}

\definecolor{green}{rgb}{0,0.5,0}

\definecolor{yellow}{rgb}{1,0.549,0}

\definecolor{red}{rgb}{0.6,0.0,0.0}

\definecolor{darkred}{rgb}{0.9,0.4,0}

\definecolor{purple}{rgb}{0.58,0,0.827}

\definecolor{backgcode}{rgb}{0.97,0.97,0.8}
\definecolor{Brown}{cmyk}{0,0.81,1,0.60}
\definecolor{OliveGreen}{cmyk}{0.64,0,0.95,0.40}
\definecolor{CadetBlue}{cmyk}{0.62,0.57,0.23,0}

\DeclareMathOperator*{\argmin}{arg\,min~}

\newcommand{\qu}[1]{``{#1}''}
\newcommand{\bv}[1]{\boldsymbol{#1}}

\newcommand{\bSigma}{\bv{\Sigma}}
\newcommand{\bSigmaw}{\bv{\Sigma}_{W}}

\newcommand{\betaT}{\beta_T}

\newcommand{\betaThat}{\hat{\beta}}

\newcommand{\ybar}{\bar{y}}

\newcommand{\ybarT}{\ybar_T}
\newcommand{\ybarC}{\ybar_C}

\newcommand{\inddist}{~{\buildrel ind \over \sim}~}

\newcommand{\half}{\frac{1}{2}}

\newcommand{\B}{\bv{B}}

\newcommand{\M}{\bv{M}}

\newcommand{\X}{\bv{X}}

\newcommand{\p}{\bv{p}}

\newcommand{\I}{\bv{I}}

\newcommand{\Y}{\bv{Y}}
\newcommand{\W}{\bv{W}}

\newcommand{\x}{\bv{x}}

\newcommand{\w}{\bv{w}}

\newcommand{\tauhat}{\hat{\tau}}

\newcommand{\onevec}{\bv{1}}
\newcommand{\zerovec}{\bv{0}}

\newcommand{\y}{\bv{y}}

\renewcommand{\u}{\bv{u}}
\renewcommand{\v}{\bv{v}}

\renewcommand{\a}{\bv{a}}

\newcommand{\bbeta}{\bv{\beta}}

\newcommand{\reals}{\mathbb{R}}

\newcommand{\eqncomment}[1]{\quad \text{(#1)}}

\newcommand{\beqn}{\vspace{-0.25cm}\begin{eqnarray*}}
\newcommand{\eeqn}{\end{eqnarray*}}
\newcommand{\bneqn}{\vspace{-0.25cm}\begin{eqnarray}}
\newcommand{\eneqn}{\end{eqnarray}}
\newcommand{\benum}{\begin{enumerate}}
\newcommand{\eenum}{\end{enumerate}}

\newcommand{\parens}[1]{\left(#1\right)}
\newcommand{\squared}[1]{\parens{#1}^2}
\newcommand{\angbrace}[1]{\left<#1\right>}

\newcommand{\prob}[1]{\mathbb{P}\parens{#1}}

\newcommand{\cprob}[2]{\prob{#1~|~#2}}

\newcommand{\bracks}[1]{\left[#1\right]}
\newcommand{\braces}[1]{\left\{#1\right\}}

\newcommand{\abss}[1]{\left|#1\right|}
\newcommand{\norm}[1]{\left|\left|#1\right|\right|}
\newcommand{\normsq}[1]{\norm{#1}^2}

\newcommand{\expe}[1]{\mathbb{E}\bracks{#1}}

\newcommand{\cexpesub}[3]{\expesub{#1}{#2\,|\,#3}}

\newcommand{\expesub}[2]{\mathbb{E}_{\,#1}\bracks{#2}}

\newcommand{\var}[1]{\mathbb{V}\text{ar}\bracks{#1}}
\newcommand{\mse}[1]{\mathbb{M}\text{SE}\bracks{#1}}
\newcommand{\varsub}[2]{\mathbb{V}\text{ar}_{#1}\bracks{#2}}
\newcommand{\cvarsub}[3]{\mathbb{V}\text{ar}_{#1}\bracks{#2\,|\,#3}}

\renewcommand{\exp}[1]{\mathrm{exp}\parens{#1}}

\newcommand{\oneover}[1]{\frac{1}{#1}}
\newcommand{\overtwo}[1]{\frac{#1}{2}}

\newcommand{\bernoulli}[1]{\mathrm{Bern}\parens{#1}}

\newcommand{\zeroonecl}{\bracks{0,1}}

\newcommand{\ourtitle}{The Role of Pairwise Matching in Experimental Design for an Incidence Outcome}
\title{
\ourtitle
}

\author[1]{Adam Kapelner\thanks{Electronic address: \texttt{kapelner@qc.cuny.edu}; Principal Corresponding author}}
\author[2]{Abba M. Krieger\thanks{Electronic address: \texttt{krieger@wharton.upenn.edu}; Corresponding author}}
\author[3]{David Azriel\thanks{Electronic address: \texttt{davidazr@technion.ac.il}; Corresponding author}}

\affil[1]{\small Department of Mathematics, Queens College, CUNY}
\affil[2]{Department of Statistics, The Wharton School of the University of Pennsylvania}
\affil[3]{Faculty of Industrial Engineering and Management, The Technion, Haifa, Israel}

\begin{document}
\maketitle

\begin{abstract}
We consider the problem of evaluating designs for a two-arm randomized experiment with an incidence (binary) outcome under a nonparametric general response model. Our two main results are that the priori pair matching design of Greevy et al. (2004) is (1) the optimal design as measured by mean squared error among all block designs which includes complete randomization. And (2), this pair-matching design is minimax, i.e. it provides the lowest mean squared error under an adversarial response model. Theoretical results are supported by simulations and clinical trial data.
\end{abstract}
\vspace{6.5cm}
\pagebreak

\section{Introduction}\label{sec:intro}

Our goal is to examine the role of experimental design when estimating a treatment effect via the difference-in-means estimator for the average treatment effect in a two-arm treatment-control randomized controlled trial (RCT). Each of the $2n$ \emph{subjects} is \emph{assigned} i.e. administered a value $w_i$ to the treatment group ($w_i := +1$) or a control group ($w_i := -1$) and all values together is called an \emph{assignment} $\w := \bracks{w_1, \ldots, w_{2n}}^\top$. Each subject's \emph{covariates} $\x_i$, a vector of real values with length $d$, are known beforehand and considered fixed. At the completion of the study, an incidence \emph{response} $\y = \bracks{y_1, \ldots, y_{2n}}^\top$ in the set $\braces{0,1}$ is collected (e.g. cardiac event vs. no cardiac event). We will assume the covariates and the treatment are related to the probability of a positive response $p_{i} := \cprob{Y_i = 1}{x_i, w_i}$ but we do not assume a functional form for this probability function. This setting of \qu{assign subjects first and assess response later} is sometimes called \qu{non-sequential} and was the classic assignment setting studied by \citet{Fisher1925} when he was assigning fertilizer treatments to agricultural plots. The setting is of great importance today. For example they occur in clinical trials as \qu{many phase I studies use `banks' of healthy volunteers ... [and] ... in most cluster randomised trials, the clusters are identified before treatment is started} \citep[page 1440]{Senn2013}.

The practitioner has a choice before the experiment begins that can affect the efficiency of the estimation: the experimental \emph{design} --- the set of \emph{allocations} drawn from when assigning the subjects to either treatment or control. This is a well-studied problem when the response is continuous, but less well-studied when not. The naive design allows for every assignment $\w \in \braces{+1,-1}^{2n}$ and is called the \qu{Bernoulli Trial} \citep[Section 4.3]{Imbens2015}. This is an unpopular design in the nonsequential setting as it allows for differing numbers of subjects in the two experimental arms. To avoid this possibility, the experimenter frequently \emph{restricts} the two arms to have the same number of subjects and this we call balanced complete randomization design (BCRD). Designs that are even more restrictive are popular in practice, e.g. blocking \citep[BL][]{Fisher1925} and rerandomization \citep{Student1938, Morgan2012}. Less popular in practice are pairwise matching \citep[PM][]{Greevy2004} and optimal single-assignment designs \citep{Bertsimas2015}. 

The fundamental question \qu{which of these myriad \emph{restricted designs} are truly optimal?} was recently answered in the case of continuous response. \citet{Kallus2018} demonstrates that the minimax variance design for the difference-in-means estimator is contingent upon the space of the response function as specified by its functional norm. Under this unified theory, the BL design emerges to be minimax under the supremum norm, the pairwise matching design emerges to be minimax under the Lipschitz norm, the rerandomization design emerges to be minimax under a linear norm and others. However, without knowledge of the space to which the response function belongs, there is \qu{no free lunch} and the classic Fisherian design of complete randomization is vindicated as the minimax strategy. To our knowledge, no systematic overarching theory has been developed in our setting of an incidence response.

Design in non-linear models, especially logistic regression has a long literature but focused on settings that are tangential to our setting. \citet{Schein2007} develop a rubric for evaluating active learning designs. These designs evaluate how the estimator performs under many possible future subjects and selects the subject and its assignment which are optimal to collect a response. \citet{Park2019, Johnson2009, Mancenido2019} and many others before them take a classic optimal design approach in response surface methodology \citep{Box1951} employing a holistic metric of the estimation of all design parameters via $D$-optimality or other classic metrics of optimality \citep{Myers1994}. The design problem in these works are different than in out setting, as they study the situation where the $\x$'s could be determined by the experimenter. Their approach (1) requires an explicit probability model for the response meaning an explicit link function $\phi$ (e.g., the expit or probit function) between the probability and the embedded function of the parameters (e.g. the linear model $\beta_0 + \bbeta^\top\x_i + \betaT w_i$). And (2) the criterions themselves frequently are recursively dependent on knowledge of the $d+2$ unknown parameters, the $\beta_j$'s. To address this situation, many authors subjectively posit values for the $d+2$ unknown parameters then proceed to create a locally optimal design based on these posited $\beta_j$'s. This procedure can be arbitrarily inaccurate if the initial $\beta_j$'s prove to be at odds with the true parameter values \citep[Section 1.1]{Abdelbasit1983}. Other authors sequentially run the experiment to minimize the chance of an egregious disparity between initial and actual parameter values (Section 1.2, ibid). And others use a Bayesian approach by specifying a prior on the $\beta_j$'s and margining their effect out using expectation (Section 1.3, ibid).

Herein, we are interested in maximizing the efficient estimation of only one parameter (the average treatment effect), do not demand an explicit $\phi$ and do not wish to rely on initial estimates of the $d+2$ parameters. In our set-up, the allocation vector $\w$ is to be determined, as apposed to the $\x$'s as in the classical setting. Furthermore, we demand the non-sequential setting where we must provide the entire $\w$ vector of individual assignments without the luxury of observing any part of $\y$. To our knowledge, our setting is relatively unexplored.

Although unexplored, our setting is widespread as there are countless clinical trials run annually where the primary outcome is an incidence metric. By way of example, \citet{Bjermer2003} measured incidence of asthma exacerbation, \citet{Julius2004} measured incidence of cardiac mortality, \citet{Zietman2010} measured incidence of failure to improve prostate outcome. Each of these examples employed the restricted design of block randomization. A cursory search of other contemporary clinical trials with primary incidence outcomes confirms this is the default design.

Herein, we demonstrate that the PM design of \citet{Greevy2004} always outperforms BCRD and BL in mean squared error and PM is the minimax design over all possible designs when the probability model is adversarial. Section~\ref{sec:background} sets up our problem, introduces the designs we consider and explains how these designs produce assignments $\w$. Section~\ref{sec:results} records our theoretical results, Section~\ref{sec:sims} provides simulation evidence of our theoretical results (including an example using clinical trial data) and Section~\ref{sec:discussion} concludes.

\section{Our Model and Designs}\label{sec:background}

Given the allocation vector $\w$, the $2n$ binary subject responses $\y$ are assumed to be realizations from an independent Bernoulli process with a subject-specific parameter conditional on the allocation, i.e. 

\bneqn\label{eq:ind_bernoulli}
Y_i \inddist \bernoulli{p_{i}}, \quad i \in \braces{1, \ldots, 2n},
\eneqn

\noindent where we denote the vector $\Y := \bracks{Y_1, \ldots, Y_{2n}}$. The subject-specific parameter is a function of the covariates $\x_i$ and treatment, $p_i(\x_i, w_i)$. For convenience we denote the vectors $\p_T := \bracks{p_1(\x_1, +1), \ldots, p_{2n}(\x_{2n}, +1)}^\top$ and $\p_C := \bracks{p_1(\x_1, -1), \ldots, p_{2n}(\x_{2n}, -1)}^\top$. We define our parameter of interest as the sample average treatment effect (SATE),

\bneqn\label{eq:tau_def}
\tau := \oneover{2n} (\p_T - \p_C)^\top \onevec.
\eneqn

\noindent which is also known as the \qu{mean risk difference} or \qu{mean rate difference}. 

The experimental design $\W$ is a multivariate shifted-and-scaled Bernoulli that produces vectors of assignments $\w \in \braces{-1,+1}^{2n}$ assumed to be uniform over its support, $\prob{\W = \w_j} = \prob{\W = \w_k}$ for all $j, k$. We make two restrictions on the designs considered. First, (A1) we ensure that all assignments produce an equal number of treatment and control assignments, i.e., $\w^\top \onevec = 0$ for all $\w$ in the support of $\W$. Second, (A2) we assume each individual subject has equal probability of being assigned to either arm, i.e., $\expe{W_i} = 0$ for all $i$. Note that (A2) is a weaker assumption than $\prob{\W = \w} = \prob{\W = -\w}$, which is common in the design literature and implies (A2).


We assume the only sources of the randomness in the responses are (R1) the treatment assignments $\w$ and (R2) the drawing of the 0/1 from the Bernoulli random variable of Equation~\ref{eq:ind_bernoulli}. (R1) is termed the \emph{randomization model} \citep[Chapter 6.3]{Rosenberger2016} whereby \qu{the $2n$ subjects are the population of interest; they are not assumed to be randomly drawn from a superpopulation} \citep[page 297]{Lin2013}. The covariates $\x_1, \ldots, \x_{2n}$ are considered fixed. 

Our nonparametric estimator for $\tau$ is 

\bneqn\label{eq:estimator}
\tauhat := \oneover{n} \W^\top \Y.
\eneqn

\noindent After $\w$ is realized from $\W$ and $\y$ is realized from $\Y$, the familiar classic difference-in-means estimate $\ybarT - \ybarC$ emerges where $\ybarT$ and $\ybarC$ denote the average of the responses in the treatment and control group respectively. 

We show in Section~\ref{app:unbiasedness_of_estimator} of the Supplementary Material that $\tauhat$ is unbiased for $\tau$ when taking the expectation over (R1) and (R2). Thus, its mean squared error (MSE) is equal to its variance, which is derived in Section~\ref{app:variance_of_estimator} of the Supplementary Material to be

\bneqn\label{eq:mse}
\mse{\tauhat} = \oneover{4 n^2}\parens{\v^\top \bSigma \v + 2(\p_T^\top (1 - \p_T) + \p_C^\top(1-\p_C))},
\eneqn

\noindent where $\v := \p_T + \p_C$ and $\bSigma := \var{\W}$, the variance-covariance matrix of all the assignments $\w$ produced by the design. Since the only term that is dependent on the design is the quadratic form $\v^\top \bSigma \v$, this quadratic form term is of special interest and will be the objective function to be compared among designs in Section~\ref{sec:results}. The other terms have a nice interpretation as the sum of the subjects'  variances of the Bernoullis for both arms' incidence responses.

As the $2n \times 2n$ variance-covariance matrix of the design plays a fundamental role, we will explain its values for the designs we consider here. For any design, the diagonal values are all 1 as $\expe{W_i}=0$  by property (A2) and $W_i^2=1$ for all $i$ since $W_i \in \{-1,+1\}$.
Thus, experimental designs differ in their degree of dependence between assignments $W_i$ and $W_j$ codified by the off-diagonal elements.  

Consider the BL design with $B$ blocks where each are equally-sized of size $n_b = 2n / B$ for all $b \in \braces{1, \ldots, B}$ subjects each. The variance-covariance matrix for BL, $\bSigma_{BL(B)}$, is a block-diagonal matrix with $B$ blocks each of size $n_b \times n_b$ as the subjects between blocks are independent. The off-diagonal entries within blocks are $-1/(n_b-1)$ because if one subject in the block is assigned to the treatment arm, this makes it a bit more probable that the other subjects in the block are assigned to the control arm as the number of treatment and control subjects must be equal within the block. The BCRD design can then be thought of as one large block and thus all off-diagonal entries in its variance-covariance matrix, $\bSigma_{BL(1)}$, are $-1/(2n-1)$. And the PM design can be thought of as the case where there are $n_b = 2$ subjects per block for all $b$ with $B = n$ total blocks and thus its variance-covariance matrix $\bSigma_{BL(n)}$ is block-diagonal with $2 \times 2$ blocks with off-diagonal entries of $-1/(2-1) = -1$. 

How are these blocks created in BL? To obtain the theoretical results of Section~\ref{sec:results}, we must assume these blocks are created with an optimal match structure $\mathcal{M}^*$ created from $\v$ \citep[see][Section 2.2.3]{Krieger2022}. The structure is formally a set of $B$ tuples of sizes $n_1, \ldots, n_B$, each set indicating the subjects indicies of the subjects belonging to each block. This optimal match structure is created by first sorting the values of $v_i$ and recording the order of sorted subject indicies. Then the $B$ tuples would fill up in order. For example if the blocksize is homogenous with all $n_b = 4$, the design $\W$ would view sorted subjects numbers 1, 2, 3, 4 as a 4-tuple and randomize with 1/6 probability between the permutations $\angbrace{+1,+1,-1,-1}$, $\angbrace{+1,-1,+1,-1}$, \ldots, $\angbrace{-1,-1,+1,+1}$. Analogously, sorted subjects numbers 5, 6, 7, 8 would be a 4-tuple and randomized in the same fashion, etc. In the PM design, the design $\W$ would view sorted subjects numbers 1, 2 as a pair and randomize with 1/2 probability between $\angbrace{+1,-1}$ and $\angbrace{-1,+1}$. Then sorted subjects numbers 3, 4 would be a pair and randomized in the same fashion, etc. Thus in PM, $\mathcal{M}^* = \braces{\angbrace{i_1^*,j_1^*}, \ldots, \angbrace{i_n^*,j_n^*}}$ whose elements specify the indicies of the $n$ optimal pairs.

How can the match structure $\mathcal{M}^*$ be created in practice if $\v$ is unknown? In the case of $d=1$ (one covariate $x$ is measured for each subject), if we assume the functional form of $p_i(x_i, w_i)$ is monotonic in $x$ then either $x_1 \leq x_2 \leq \ldots \leq x_{2n}$ implies $v_1 \leq v_2 \leq \ldots \leq v_{2n}$ or $v_{2n} \leq v_{2n -1} \leq \ldots \leq v_1$. In this case ordering the subjects by their covariate value will sort subjects by their $v_i$ values.  This monotonicity is standard for instance when the probability is assumed to be a function with a linear term in $x$ embedded in a monotonic link function with range $(0,1)$ e.g. 

\bneqn\label{eq:prob_model}
p_i(x_i, w_i) := \phi(\beta_0 + \beta_1 x_i + \betaT w_i)
\eneqn

\noindent where $\phi$ could be the popular functions \texttt{expit}, \texttt{probit}, \texttt{inverse-cloglog} or any other inverse CDF of a continuous function with support on all real numbers. 

What if this monotonic assumption cannot be assumed? For example, in the case where the embedded function in Equation~\ref{eq:prob_model} has a quadratic term in the one covariate, then ordering by the $x_i$'s will not order the subjects by their $v_i$ values. However, sorting by the $v_i$ values is necessary only to create the optimal pairwise match structure $\mathcal{M}^*$. An approximate pairwise match structure $\mathcal{M}$, albeit suboptimal, would likely still perform well. We explore this setting in the simulations of Section~\ref{sec:sims} . 

If $d>1$, knowledge of the $v_i$'s is equivalent to knowledge of the full functional form of $p_i(\x_i, w_i)$. For instance, in the generalized linear model $\phi(\beta_0 + \bbeta^\top \x_i + \betaT w_i)$, one would need to know the values of the parameters $\beta_0, \bbeta$ and $\betaT$. These parameters are unknown and we find ourselves ironically in the setting of those who design experiments using $D$-optimality as we wrote about above in Section~\ref{sec:intro}. Once again, in practice, an approximate match structure $\mathcal{M}$ would likely still perform well. We explain how we obtain this approximate match structure and explore the MSE performance of these designs in the case of more than one covariate in Section~\ref{sec:sims}.

\section{Theoretical Results}\label{sec:results}


For the first two results, assume the subjects are sorted by their unknown values of $v_i$. We first prove in Section~\ref{app:bcrd_sucks} of the Supplementary Material the following result. \\

\begin{theorem}[For any sample size, PM is optimal among all block designs, including BCRD]\label{thm:block_monotonicity}
Under the model of Equation \ref{eq:ind_bernoulli}, the MSE for PM, the block design with $n$ blocks, is lower than the MSE for any block design with less than $n$ blocks where the blocksize is even. Thus, for $B < n$ and all blocksizes even but not necessarily equally-sized, $\mse{\tauhat_{PM}} < \mse{\tauhat_{BL(B)}}$ for any $\v$, where $\tauhat_{PM}$, $\tauhat_{BL(B)}$ are the estimator of Equation~\ref{eq:estimator} under matching and the block design with $B$ blocks, respectively.\\
\end{theorem}

Block designs do not span the space of all possible experimental designs. We believe it is impossible to solve for the \qu{optimal design} among the entire space of designs, i.e., to compute the measure corresponding to $\W_* := \argmin_{\W \in \mathcal{W}} \braces{\mse{\tauhat_{\W}}}$ where $\mathcal{W}$ denotes the space of all designs that satisfy assumptions (A1) and (A2) from Section~\ref{sec:background}. First of all, the MSE is a function of the design only through $\bSigma$. So at best, we could theoretically find the variance-covariance matrix of the optimal design, i.e., $\bSigma_* := \argmin_{\bSigma \in \mathcal{S}} \braces{\v^\top \bSigma \v}$ where $\mathcal{S}$ denotes the space of all variance-covariance matrices of a multivariate-Bernoulli whose realizations satisify assumptions (A1) and (A2) from Section~\ref{sec:background}. Once $\bSigma_*$ is located, it corresponds to very many different equally optimal designs $\braces{\W_*}$ as the multivariate Bernoulli random variable model has $2^{2n} - 1$ parameters with non-unique second moments \citep[Section 2.3]{Teugels1990}. Further, the optimal design would be conditional on the unknown value of $\v$.

Instead, we prove what is tractable: a theorem about the minimax design. In Section~\ref{app:pm_minimax} in the Supplementary Material we demonstrate the following result.\\

\begin{theorem}[PM is minimax]\label{thm:pm_minimax}
Under the model of Equation \ref{eq:ind_bernoulli},

\beqn
\max_{\v \in \mathcal{V}} \braces{\mse{\tauhat_{PM}}} = \min_{\W \in \mathcal{W}} \max_{\v \in \mathcal{V}} \braces{\mse{\tauhat_{\W}}},
\eeqn

\noindent where $\tauhat_{\W}$ denotes the estimator of Equation~\ref{eq:estimator} under a an arbitrary design $\W$, $\mathcal{W}$ denotes the space of all designs that satisfy assumptions (A1) and (A2) from Section~\ref{sec:background} and $\mathcal{V}$ is the space of all sorted vectors $\v := \p_T + \p_C$, the sum of the two probability parameter vectors, $\mathcal{V} := \braces{\v~:~0 \leq v_1 \leq v_2 \leq \ldots \leq v_{2n} \leq 2}$.\\
\end{theorem}

If the subjects are instead randomly sorted with respect to their $v_i$ values, we have the following robustness result for PM proved in Section~\ref{app:pm_robust} of the Supplementary Material.

\begin{remark}[PM is robust to suboptimal matching]\label{rem:pm_robust}
If the matches are randomly assigned then the MSE of PM is the same as BCRD.
\end{remark}

\noindent There is a limit to this robustness. If the matches are adversarial, then PM can perform worse than BCRD. Remark \ref{rem:pm_bad} below states a sufficient and necessary conditions for a general PM to have lower MSE than BCRD, and is 
proved in Section~\ref{app:pm_not_robust} of the Supplementary Material.

\begin{remark}[Sufficient and necessary condition for PM to do worse than BCRD]\label{rem:pm_bad}
We have that 
\[
\mse{\hat{\tau}_{BCRD}}-\mse{\hat{\tau}_{PM}}=\frac{1}{4n^2} \parens{ \frac{1}{2n-1} \sum_{i<j} (v_i - v_j)^2 -\sum_{k=1}^n (v_{i_k}- v_{j_k})^2  },
\]
where the pairs $\angbrace{i_k,j_k}$, $k=1,\ldots,n$ are in the match set $\mathcal{M}$.
Therefore, BCRD outperforms PM iff
\begin{equation}\label{eq:cond}
 \frac{1}{n(2n-1)} \sum_{i<j} (v_i - v_j)^2  < \frac{1}{n} \sum_{k=1}^n (v_{i_k}- v_{j_k})^2.
\end{equation} 
\end{remark}
To interpret the condition in Equation~\ref{eq:cond}, notice that the right-hand side is the average squared distance over all pairs $v_i,v_j$ with $i<j$ and the right-hand side is the average over the pairs in the match set ${\cal M}$. It follows that BCRD yields lower MSE if the pairs in the match set are more distant than an average pair, and otherwise PM is better.

Comparing BCRD to PM, as in Remark \ref{rem:pm_bad} is equivalent to quantifying the quality of the match through the R-squared one gets in the analysis of variance where the covariate is the $n$ level categorical variable denoting the match set. The total sum of squares is $\sum_{i=1}^{2n} (v_i - \bar{v})^2$ and the error sum of squares 
is
\[
\sum_{k=1}^n \left(v_{i_k}- \frac{v_{i_k}+v_{j_k}}{2} \right)^2+\sum_{k=1}^n \left(v_{j_k}- \frac{v_{i_k}+v_{j_k}}{2} \right)^2=\frac{1}{2} \sum_{k=1}^n (v_{i_k}- v_{j_k})^2.
\]
Therefore, the value of  R-squared is
\begin{equation}\label{eq:R_PM}
1-\frac{\frac{1}{2} \sum_{k=1}^n (v_{i_k}- v_{j_k})^2}{\sum_{i=1}^{2n} (v_i - \bar{v})^2}.
\end{equation}

By Remark \ref{rem:pm_robust}, BCRD is equivalent to creating a match set randomly. The expected value of this R-squared for BCRD is, by Equation~\ref{eq:n_b},
\begin{equation}\label{eq:R_BCRD}
1-\frac{\frac{1}{2} \frac{2n}{2n-1}\sum_{i=1}^{2n} (v_i - \bar{v})^2}{\sum_{i=1}^{2n} (v_i - \bar{v})^2}=\frac{n-1}{2n-1},
\end{equation}
i.e., it does not depend on $\v$ and increases to 1/2 as $n$ increases. Comparing Equations~\ref{eq:R_PM} and \ref{eq:R_BCRD} is equivalent to the condition in Equation \ref{eq:cond}. 

 It is possible for PM to perform better or worse than BCRD depending on the quality of the matches. The ideal match pairs the smallest $\v$ with the second smallest $\v$, the third smallest $\v$ with the fourth smallest $\v$ and so on. The worst possible match tends to put the largest $\v$ with the smallest $\v$, the second largest $\v$ with the second smallest $\v$ and so on. It is easy to see what occurs if the values of $\v$ are equally spaced. In this case, the R-squared for PM can vary anywhere from 0 to 1 as compared to an R-squared near 1/2 for BCRD. 

Further, if the covariates are not related to the response, all designs are equally performant as proven in Section~\ref{app:all_designs_equal}.

\begin{remark}[All Designs are Equal if Covariates are Uninformative]\label{rem:equality}

If $\v$ is constant, the MSE of any experimental design is $\oneover{4 n^2}\parens{2(\p_T^\top (1 - \p_T) + \p_C^\top(1-\p_C))}$.
\end{remark}


%
%
%
%

\section{Simulation Results}\label{sec:sims}

\subsection{Simulated Data}\label{sec:sims_fake}

We begin by simulating the case of one covariate $x$ measured per subject ($d=1$). We employ the model of Equation~\ref{eq:prob_model} using $\phi = $ \texttt{expit} corresponding to the classic logistic regression case of no model misspecification,

\bneqn\label{eq:fake_sim_model}
p_i(x_i, w_i) := \frac{\exp{\beta_0 + \beta_1 x_i + \betaT w_i}}{1 + \exp{\beta_0 + \beta_1 x_i + \betaT w_i}}.
\eneqn

\noindent In this case, we are guaranteed to have the optimal match structure $\mathcal{M}^*$ by sorting the $x_i$ values since the probability function is monotonic in $x$ (see discussion in Section~\ref{sec:background}). This also provides optimal block structure for any number of blocks.

We generate the values of $x$ using the standard logistic distribution quantiles evenly spaced between 0.005 and 0.995 with spacing varying by each sample size which we set to be $2n \in \braces{64, 128, 256}$. Stacking the $x_i$'s rowwise gives us the one-column matrix $\X$. We set the response parameters to be $\beta_0 = 4$, $\beta_1 = 2$ and $\betaT = 1$. Given $\X$ and these $\beta$'s, we can compute $\p_T$ and $\p_C$ which allows us to compute the risk difference, our main parameter of interest $\tau$ of Equation~\ref{eq:tau_def} precisely. The $\p_T$ and $\p_C$ values (selected via setting the $\beta$'s) attempt to de-emphasize the role of the component of the MSE independent of the design (see Equation~\ref{eq:mse}) in order that the estimated design differences will be more salient. 

For the BL design, we employ $B=8$ blocks for all sample sizes. We generate $N_{sim} = 1,000,000$ assignments $\w$ from each design for each sample size. For each $\w$, we generate all the $p_i(x_i, w_i)$'s which allows us to draw the random responses $y_i$'s from independent Bernoulli realizations (Equation~\ref{eq:ind_bernoulli}). Using the responses and the assignment we can then compute the risk difference estimate, log odds ratio and the estimate of $\betaT$ from a logistic regression (which is the only estimate to additionally require $\X$). The first estimate (of $\tauhat$ of Equation~\ref{eq:estimator}) was the subject of our theoretical investigation herein. We include the other two estimates in this simulation only to provide intuition about their theoretical performance which we leave to future work. (Note that the true log odds ratio parameter does not need computation as it is $\betaT$ and independent of $\X$ by construction in our data generating process of Equation~\ref{eq:fake_sim_model}). We then average all estimates over the $N_{sim}$ replicates to generate average estimates. We use the average estimates to compute the estimated mean squared error using the true parameter values explained previously. 

\begin{figure}[ht]
\centering
\includegraphics[width=6.6in]{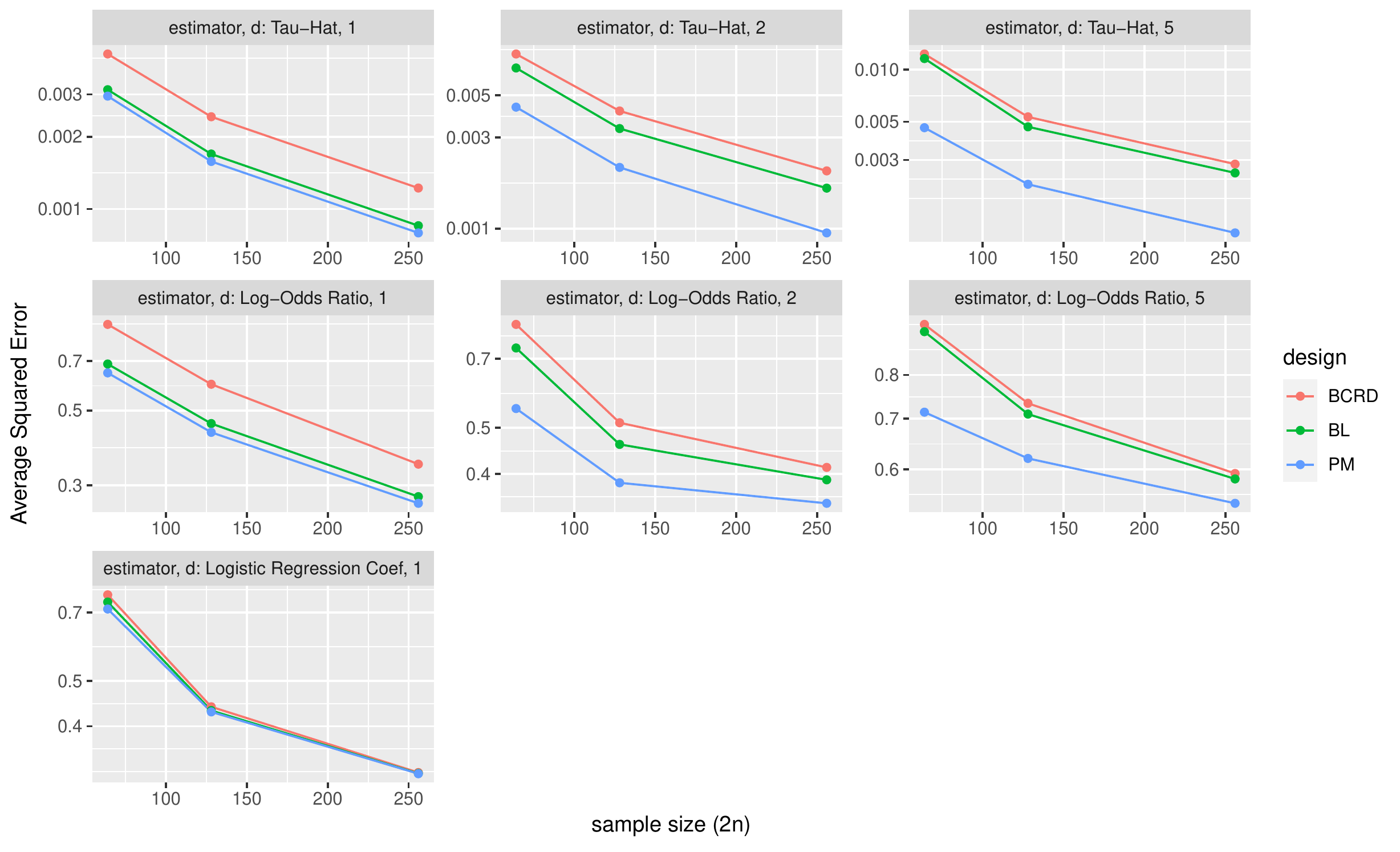}
\caption{Performance results on the simulated data for all sample sizes, number of covariates, estimates and the three designs BCRD, BL and PM. The top row show the average estimates for $\tauhat$, the second row show the average estimates for the log odds ratio and the last row shows the average treatment coefficient estimates in the logistic regression.}
\label{fig:fake_data_results}
\end{figure}

These mean squared error estimates for $d=1$ appear in the left-most column of Figure~\ref{fig:fake_data_results}. We can see that the PM design barely outperforms the BL design. But both designs outperform BCRD by a large margin. This result is expected by Theorem~\ref{thm:block_monotonicity}. This performance also extends to the log odds ratio estimator and the logistic regression estimator which were not theoretically explored. Additionally, for the logistic regression estimate results (lower left), we observe that BCRD is not at an extreme disadvantage. This is likely due to the fact that logistic regression employs a posteriori adjustments for covariate imbalance which reduces estimation error. For example, in the setting of continuous responses when comparing the OLS estimator for continuous response to the classic difference-in-mean estimator, the estimation error of the former is an entire order in $n$ smaller than the latter \citep[see][Equations 7 and 14]{Kapelner2021}.

The case of $d=1$ is unrealistic as real clinical trials have more than one characteristic measured per subject. To understand design performance in this more realistic case, we now simulate $d = 2$ and $d = 5$. We employ the same $\beta$'s but duplicate the previous $\beta_1$ coefficient for all $d$ covariates, i.e., $\beta_0 = 4$, $\bbeta = 2 \onevec_d$ and $\betaT = 1$. We will discuss how $\X$ is generated in the coming paragraphs. We first must discuss the problem of how to create the blocks for BL design and the pairs for the PM design when $d>2$. In $d=1$, the order of the $v_i$ elements correspond to the order of the $x_i$ elements but in $d>1$, this is not the case, so we need to use the information about the $\x_i$'s to approximate the order of the $v_i$'s. Thus the theorems of Section~\ref{sec:results} do not apply when $d>2$ in a strict sense. The simulations herein provide intuition about the theoretical performance in the case of the BL and PM designs when the blocks are imperfectly constructed.

For the PM design, we must generate $n$ pair matches. To do so, we employ the optimal nonbipartite matching algorithm \citep[see][]{Lu2011} using the R package \texttt{nbpMatching} \citep{Beck2016}. This algorithm requires a specified distance function between two subjects' covariate vectors, $\x_i$ and $\x_j$, to generate a distance matrix and then remarkably solves the minimum sum of all pair distances problem in polynomial time. We employ the Mahalanobis distance which was recommended by \citet{Rubin1979}, the first work that demonstrated the robustness of matching in regression with a non-linear response model. As this algorithm returns the same matches for a distance function scaled by a multiplicative constant, we employ the proportional between-subjects Mahalanobis distance, $(\x_i - \x_j)^\top \hat{\Sigma}_{\X}^{-1} (\x_i - \x_j)$ where $\hat{\Sigma}_{\X}^{-1}$ is the $d \times d$ sample variance-covariance matrix of all $2n$ subjects' $d$ covariate vectors \citep[Section 2.2]{Stuart2010}.

For the BL design, we wish to retain $B=8$ blocks for all sample sizes and number of covariates. For $d=2$ we use four blocks for the first covariate and two blocks for the second. For $d=5$, we use two blocks for each of the first three covariate and do not block on the remaining covariates. This is standard in practice; experimenters block on a subset of covariates that are a priori conjectured to have the most pronounced effect on the response. There is an additional problem: blocks created from continuous covariate data will likely be heterogeneously-sized (i.e., $n_b$ will vary block-block) and the sizes may be uneven. Since our theoretical results are proven for even-sized blocks, one way to enforce this setting is to enforce homogeneity. To do so, we generate the homogeneous block designs first and then populate the $\X$ matrix after. To do so, the first covariate is always generated in order of the quantiles of the standard logistic distribution as explained above for the $d=1$ case. For the $d=2$ case, the second covariate is generated as well via the quantiles of the standard logistic distribution, shuffled, spliced in half, ordered, spliced in half again to create the four blocks, then reshuffled. An analogous procedure is followed for the third covariate when $d=5$.

Performance results can be seen in columns 2 and 3 of Figure~\ref{fig:fake_data_results}. We mostly observe the same results as the $d=1$ setting except the dominance of PM is more pronounced. Note that we omit the results for the logistic regression estimates here which were unstable.

\subsection{Clinical Data}\label{sec:sims_real}

In this section, we use clinical trial data from \citet{Foster2010}, a 12-week, multicenter, double-blind, placebo-controlled sequential RCT investigating whether amitriptyline, an antidepressant drug, can effectively treat painful bladder syndrome. The endpoint we chose was the final pain reading after 12 weeks. Since this was a continuous endpoint, we coerced it to incidence by thresholding at the value of the sample median. The original RCT employed a Bernoulli Trial design \citep[Chapter 4.2]{Imbens2015} where each subject received the treatment independent with probability 50\%. The 18 covariates assessed for each patient in the study (and their data types) were: \texttt{age} (continuous), \texttt{gender} (binary), \texttt{presence of sexually transmitted disease} (binary), \texttt{lives with a partner} (binary), \texttt{baseline urinary tract infection} (binary), \texttt{patient's race is white} (binary), \texttt{patient's race is Hispanic} (binary), \texttt{level of education} (ordinal with five levels), \texttt{level of employment} (ordinal with five levels), \texttt{baseline symptom index} (continuous), \texttt{baseline anxiety and depression} (continuous), \texttt{baseline mental quality of life} (continuous), \texttt{baseline problems index} (continuous), \texttt{baseline symptom inventory} (continuous), \texttt{baseline urination freqency} (continuous), \texttt{baseline bladder-associated pain} (continuous) and \texttt{baseline bladder urgency} (continuous). The study was negative; the investigators found no statistically significant effect of the treatment over the placebo.

We sought to compare the performance of the different designs we considered as in Section~\ref{sec:sims_fake}. One way to do this comparison is via a parametric bootstrap resampling procedure. We sample $\w$'s from the designs and then $\y$'s from a probability model such as a logistic regression fitted to the entire dataset and the different $\w$'s. We calculate our estimators and compare. This simulation would lack verisimilitude if this probability model was wrong but given that we cannot perform new prospective studies under different designs, it is a good compromise for this work.

After dropping any patients with missing covariates or responses, the total number of subjects considered was $2n = 224$ where 116 subjects were administered the treatment, 108 were administered the placebo. Due to the endpoint's discreteness, the thresholding on the sample median did not split the data evenly; 132 subjects' responses were coded as zero and 92 responses were coded as 1. 

As in the previous section, we simulate $d \in \braces{1,2,5}$. We first fit a stepwise logistic regression to all 18 variables to find the importance order of the patient characteristics. In order of importance, the most important variable were \texttt{baseline symptom inventory}, \texttt{baseline mental quality of life}, \texttt{baseline symptom inventory}, \texttt{baseline bladder-associated pain} and \texttt{lives with a partner}. A regression on the most important variable yields high statistical significance of the first covariate, a regression the top $d=2$ yields high statistical significance for the first covariate and borderline statistical significance for the second covariate and a regression on the top $d=5$ yields high statistical significance for the first covariate and borderline statistical significance for the fifth covariate. Over 1,000 random training-test splits, the out-of-sample area under the curve values for each model are 0.68, 0.69 and 0.70 respectively indicating poor discrimination of the response \citep[page 177]{Hosmer2013}.

\begin{figure}[ht]
\centering
\includegraphics[width=6.6in]{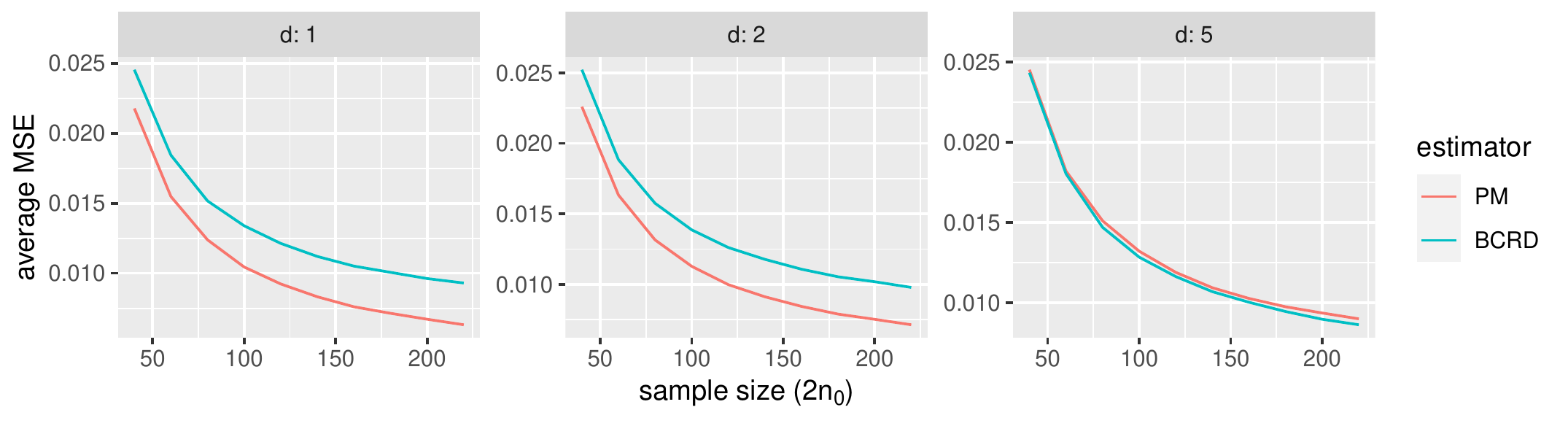}
\caption{Sample average MSE of $\betaThat$ under the PM design and the BCRD design as a function of sample size for all settings of $d$. Error bars are unavailable as simulation estimates are highly dependent on each other as they are functions of much of the same data.}
\label{fig:clinical_data_results}
\end{figure}

We use these three regression models as probability models in this parametric bootstrap study. We also add a treatment effect to the models of $\beta_T = 1$ only to induce greater separation between the two designs. Because we assume a probability model, we precompute the true value of $\beta_T$ which we seek to estimate. To understand the sample-size dependence of performance, we simulate under $2n_0 \in \braces{40, 60, \ldots, 200, 220}$. For a specific sample size, we draw a $\w$ under PM, a $\w$ under BCRD, then we draw a $\y$ under PM's allocation and finally a $\y$ under BCRD's allocation. We then compute $\betaThat$. We repeat this process approximately 100,000 times for each $n_0, d$ for a total of approximately 3,000,000 faux clinical studies. MSE results were then aggregated at the values of $n_0$ and $d$ to average the squared error results and displayed in Figure~\ref{fig:clinical_data_results}. 

The $d=1$ and $d=2$ cases are expected given our theory and the previous section's simulation results. Even though the covariates are mostly uninformative, they are informative enough to the extent that Remark~\ref{rem:equality} does not apply. The matching $\mathcal{M}$ seems to approximately the optimal matching well: the efficiency ratio of the BCRD estimator to the PM estimator at a sample size of 200 is approximately 2. This implies that PM can save 50\% of the sample size in a trial relative to BCRD. However, at $d=5$, we observe no performance edge of PM over BCRD. This is likely due to the matches being poor as there are now three of the five covariates completely uninformative and only one that is informative. Hence, we are likely matching randomly and observing no gain in PM over BCRD (Remark~\ref{rem:pm_robust}).

\section{Discussion}\label{sec:discussion}

The PM design occupies a central role when estimating a treatment's risk difference in incidence response models as demonstrated by both theoretical analysis and multiple simulations. This result is not surprising as PM's role in providing robust estimation under a nonlinear response model has a long literature in observational studies \citep{Stuart2010}. Also, PM was shown to be the minimax design in the case of continuous response when the response model belongs to a space of Lipschitz functions \citep[Section 2.3.2]{Kallus2018}, which is clearly true of popular probability models such as \texttt{expit}, \texttt{probit}, and \texttt{inverse-cloglog}. PM being minimax differs with the analogous result for continuous response models as our theoretical results apply to probability response models of a general form (not only Lipschitz functions). BCRD is the minimax design in the case of complete ignorance about the response function in the continuous setting \citep[Section 2.1]{Kallus2018}.

We make explicit our design recommendation of PM. For the case of $d=1$, one first sorts the subjects in order of the value of the one measured covariate (or composite risk metric which may be common in many clinical settings). For $d>1$, our theoretical conclusion requires sorting the subject by order of the unknown $v_i$'s which is impossible in practice. However, the simulations of Section~\ref{sec:sims} show that nonbipartite pair matching using the Mahalanobis distance metric of the covariate vector pairs can approximate matching on the unknown $v_i$'s. The intuitive reason why this performs well is that for any continuous function, $\x_i \approx \x_j$ implies that $f(\x_i) \approx f(\x_j)$. This approximation should be especially good in the usually-assumed response model of $f(\x) = \phi(\beta_0 + \bbeta^\top\x)$ with $\phi$ being a link function with a slow-moving gradient throughout most of the input space. Section~\ref{sec:sims_real} demonstrates that even in a typical clinical setting where covariates are not too relevant to a noisy endpoint, the PM design still provides improvements in estimator efficiency. This efficiency persists only for matching on a few previously-known  influential variables. As these number of variables increases, and the variables are less important, PM loses its performance edge over BCRD as the matches become more and more random relative to the underlying true $\v$ as expected by Remark~\ref{rem:pm_robust}. We do not observe evidence of adversarial matches (worse-than-random matches) in our simulations which would yield worse performance than BCRD (as elucidated in Remark~\ref{rem:pm_bad}).

There are many extensions of this work. Most glaring, we only did a theoretical analysis of the difference-in-means estimator for the risk difference parameter. We could theoretically explore the performance of the other many common estimators for risk ratio, odds ratio, log odds ratio (with and without covariate-adjustment post-assignment using a logistic regression for instance). Intuitively, we believe there are results analogous to Theorem~\ref{thm:block_monotonicity} in the settings of the nonparametric log odds ratio estimator and the covariate-adjusted log odds ratio estimator (the logistic regression coefficient) as we observed empirically (see Figures~\ref{fig:fake_data_results} and \ref{fig:clinical_data_results}) that the PM design outperforms BL which outperforms BCRD in estimator error. Further, we can investigate paired estimators for all estimators mentioned above and our difference-in-means estimator. Even in our setting of the difference-in-means estimator for risk difference, there is more work to be done: we would like to establish theory for the case of $d>1$ under the practice recommendation above of optimal nonbipartite matching via the Mahalanobis distance metric. This is important as we know that getting the matches \qu{wrong} results in performance worse than BCRD (in unshown simulations) and we would like to have a measure of risk to know when to revert to BCRD. Lastly, exploring the performance of the pairing on-the-fly method of \citet{Kapelner2021} in the sequential allocation setting is important as the vast majority of clinical trials are sequential.

\subsection*{Acknowledgments}

This research was supported by Grant No 2018112 from the United States-Israel Binational Science Foundation (BSF).

\bibliographystyle{apa}\bibliography{refs}
\pagebreak

\appendix

\begin{center}
\LARGE{Supplementary Material} \\~\\
\large{for \qu{\ourtitle} by Adam Kapelner, Abba Krieger and David Azriel}
\end{center}

\section{Proofs}\label{app:proofs}

\subsection{The Unbiasedness of the Estimator}\label{app:unbiasedness_of_estimator}

We take the expectation of the estimator $\tauhat$ (Equation~\ref{eq:estimator} of the main text) over both sources of randomness ($\Y$ and $\W$) using the law of iterated expectation. Letting $\bv{\pi} := \cexpesub{\Y}{\Y}{\w}$,

\bneqn\label{eq:iter_expe}
\expesub{\W}{\cexpesub{\Y}{\tauhat}{\w}} &=& \oneover{n} \expesub{\W}{\W^\top \cexpesub{\Y}{\Y}{\w}} = \oneover{n} \expesub{\W}{ \W^\top \bv{\pi}} \nonumber \\
&=& \oneover{n} \sum_{i=1}^{2n} \expesub{W_i}{W_i \pi_i} 
\eneqn

\noindent where the entries of the vector $\bv{\pi}$ can be written as

\bneqn\label{eq:piis}
\pi_i(w_i) := \begin{cases}
p_{T,i} ~~\text{if}~~ w_i = +1 \\
p_{C,i} ~~\text{if}~~ w_i = -1 \\
\end{cases} = \overtwo{(1 + w_i)p_{T,i} + (1 - w_i)p_{C,i}} = \overtwo{v_i + w_i (p_{T,i} - p_{C,i})}
\eneqn

\noindent where $p_{T,i} := \cprob{Y_i = 1}{\x_i, w_i = +1}$, $p_{C,i} := \cprob{Y_i = 1}{\x_i, w_i = -1}$ and $v_i := p_{T,i} + p_{C,i}$. Continuing with the iterated expectation we find that

\beqn
\expesub{\W}{\cexpesub{\Y}{\tauhat}{\w}} &=& \oneover{n} \sum_{i=1}^{2n} \expesub{W_i}{W_i \parens{\overtwo{v_i + W_i (p_{T,i} - p_{C,i})}}} \\
&=& \oneover{2n} \sum_{i=1}^{2n}  \expesub{W_i}{v_i W_i + W_i^2 (p_{T,i} - p_{C,i})} \\
&=& \oneover{2n} \sum_{i=1}^{2n}  v_i\expesub{W_i}{W_i} + (p_{T,i} - p_{C,i}) \expesub{W_i}{W_i^2}  \\
&=& \oneover{2n} \sum_{i=1}^{2n}  (p_{T,i} - p_{C,i})  \eqncomment{by (A2) and definition of $W$}\\
&=& \oneover{2n}  (\p_T - \p_C)^\top \onevec 
\eeqn

\noindent which is the definition of $\tau$ in Equation~\ref{eq:tau_def} of the main text. ~~$\blacksquare$

\subsection{The Variance of our Estimator}\label{app:variance_of_estimator}

We use the law of total variance to find the variance of $\tauhat$,

\bneqn\label{eq:total_variance}
\var{\tauhat} = \expesub{\W}{\cvarsub{\Y}{\tauhat}{\w}} + \varsub{\W}{\cexpesub{\Y}{\tauhat}{\w}}.
\eneqn

\noindent We begin computing the variance by examining the second term,, 

\bneqn\label{eq:total_variance_1}
\varsub{\W}{\cexpesub{\Y}{\tauhat}{\w}} &=& \varsub{\W}{\oneover{n}\W^\top \bv{\pi}} \nonumber \eqncomment{via Equation~\ref{eq:iter_expe}}\\
&=& \oneover{n^2}\varsub{\W}{\sum_{i=1}^{2n} W_i \parens{\overtwo{v_i + W_i (p_{T,i} - p_{C,i})}}} \nonumber \eqncomment{via Equation~\ref{eq:piis}} \\
&=& \oneover{4n^2} \varsub{\W}{\sum_{i=1}^{2n} v_i  W_i + (p_{T,i} - p_{C,i}) W_i^2} \nonumber \\
&=& \oneover{4n^2} \varsub{\W}{\v^\top \W + (\p_T - \p_C)^\top \onevec} \nonumber \\
&=& \oneover{4n^2} \v^\top \bSigma \v
\eneqn

\noindent where $\bSigma$ is the variance-covariance matrix of the random variable $\W$ which is explained in detail in Section~\ref{sec:background} of the main text. 

We now examine the term whose expectation is being taken in the first term of the total variance (Equation~\ref{eq:total_variance}),

\beqn
\cvarsub{\Y}{\tauhat}{\w} &=& \cvarsub{\Y}{\oneover{n} \w^\top \Y}{\w} \\
&=& \oneover{n^2} \sum_{i=1}^{2n} \cvarsub{Y_i}{w_i Y_i}{w_i} \\
&=& \oneover{n^2} \sum_{i=1}^{2n} w_i^2 \cvarsub{Y_i}{Y_i}{w_i} \\
&=& \oneover{n^2} \sum_{i=1}^{2n} \pi_i (1-\pi_i) \\
&=& \oneover{n^2} \parens{\sum_{i=1}^{2n} \pi_i - \pi_i^2}
\eeqn

\noindent Substituting Equation~\ref{eq:piis} for $\pi_i$ yields

\beqn
\cvarsub{\Y}{\tauhat}{\w} &=& \oneover{n^2} \sum_{i=1}^{2n} \oneover{2}\parens{v_i + w_i (p_{T,i} - p_{C,i})} - \squared{\oneover{2}\parens{v_i + w_i (p_{T,i} - p_{C,i})}} \\
&=& \oneover{2 n^2} \sum_{i=1}^{2n} v_i + w_i (p_{T,i} - p_{C,i}) - \half v_i^2 - \half w_i^2 (p_{T,i} - p_{C,i})^2 + v_i w_i (p_{T,i} - p_{C,i}) \\
&=& \oneover{2 n^2}\parens{ \v^\top \onevec + \w^\top (\p_T - \p_C) - \half \normsq{\v}  - \half \normsq{\p_T - \p_C} + \sum_{i=1}^{2n} w_i v_i (p_{T,i} - p_{C,i}}.
\eeqn

\noindent When we take the expectation over $\W$, by (A2) the terms above that depend on $w_i$ disappear:

\bneqn\label{eq:total_variance_2}
\expesub{\W}{\cvarsub{\Y}{\tauhat}{\w}} = \oneover{4 n^2}\parens{2 \v^\top \onevec - \normsq{\v} - \normsq{\p_T - \p_C}}.
\eneqn

\noindent Putting together the two terms (Equation~\ref{eq:total_variance_1} and Equation~\ref{eq:total_variance_2}) and simplifying gives us the final expression,

\beqn
\var{\tauhat} &=& \oneover{4 n^2}\parens{\v^\top \bSigma \v + 2 \v^\top \onevec - \normsq{\v} - \normsq{\p_T - \p_C}} \\
&=& \oneover{4 n^2}\parens{\v^\top \bSigma \v + \sum_{i=1}^{2n} 2 v_i  - v_i^2 - (p_{T,i} - p_{C,i})^2} \\
&=& \oneover{4 n^2}\parens{\v^\top \bSigma \v + \sum_{i=1}^{2n} 2 p_{T,i} + 2 p_{C,i}  - (p_{T,i} + p_{C,i})^2 - (p_{T,i} - p_{C,i})^2} \\
&=& \oneover{4 n^2}\parens{\v^\top \bSigma \v + \sum_{i=1}^{2n} 2 p_{T,i} + 2 p_{C,i}  - p_{T,i}^2 - p_{C,i}^2 - 2p_{T,i}p_{C,i} - p_{T,i}^2 - p_{C,i}^2 + 2p_{T,i}p_{C,i}} \\
&=& \oneover{4 n^2}\parens{\v^\top \bSigma \v + 2\sum_{i=1}^{2n} p_{T,i} + p_{C,i}  - p_{T,i}^2 - p_{C,i}^2 } \\
&=& \oneover{4 n^2}\parens{\v^\top \bSigma \v + 2(\p_T^\top (1 - \p_T) + \p_C^\top(1-\p_C))}.~~~\blacksquare
\eeqn


\subsection{PM Outperforms BL for with Even Block Sizes}\label{app:bcrd_sucks}

We compare the MSE of our estimator a BL design (with $B < n$ an block sizes $n_{1} + n_2 + \ldots + n_B = 2n$ where all $n_b$ are even) to a PM design (also a block design where $B = n$ and all $n_b = 2$). As the only term that is affected by design is the quadratic form $\v^\top \bSigma \v$ (see Equation~\ref{eq:mse} of the main text), we compare the two designs on this term. The variance-covariance matrices for the block design and the PM design denoted $\bSigma_{BL(B)}$ and $\bSigma_{PM}$ respectively are both block diagonal where $\bSigma_{BL(B)}$ has blocks $\B_b$ where $b \in \braces{1, \ldots, B}$ and $\bSigma_{PM}$ has blocks $\M_b$ where $b \in \braces{1, \ldots, n}$ and

\beqn
\B_b &:=& \bracks{\begin{array}{cccc} 
1 & -\oneover{n_b - 1} & \ldots & -\oneover{n_b - 1} \\
-\oneover{n_b - 1} & 1 & \ldots & -\oneover{n_b - 1} \\
\vdots & & \ddots & \vdots \\
-\oneover{n_b - 1} & -\oneover{n_b - 1} & \ldots & 1
\end{array}} = \frac{n_b}{n_b - 1}\I - \frac{1}{n_b - 1}\onevec\onevec^\top, \\
\M_b &:=& \bracks{\begin{array}{cc} 
1 & -1 \\
-1 & 1
\end{array}} = 2\I - \onevec\onevec^\top 
\eeqn

\noindent for all $b$. We let $\v_b$ be the subvector of $\v$ with entries equal to the entries of block $b$ relative to the entire $\bSigma_{PM}$. And we let $\a_b := \v_b -  \onevec \v_b^\top \onevec / 2$, i.e., $\a_b$ is the mean centered $\v_b$. Note that $\bSigma \onevec = \onevec^\top \bSigma = \zerovec$ for all block designs and $\a_b^\top \onevec = 0$ for all mean-centered vectors. We also make use of the following fact:

\bneqn\label{eq:vector_fact}
\normsq{\u} = \oneover{d} \sum_{1 \leq i < j \leq d} (u_i - u_j)^2~~\text{where}~~ \u \in \reals^d~\text{and}~ \u^\top \onevec = 0.
\eneqn

\noindent Using these definitions and facts we can now compute the quadratic form for the PM design. 

\bneqn\label{eq:PM_mse}
\v^\top \bSigma_{PM} \v &=& \sum_{b=1}^n \v_b^\top \M_b \v_b \nonumber \\
&=& \sum_{b=1}^n \parens{\a_b +  \onevec \half \v_b^\top \onevec}^\top \M_b \parens{\a_b + \onevec \half\v_b^\top \onevec} \nonumber \\
&=& \sum_{b=1}^n \a_b^\top \M_b \a_b \nonumber \\
&=& \sum_{b=1}^n \a_b^\top (2\I - \onevec\onevec^\top ) \a_b \nonumber \\
&=& 2 \sum_{b=1}^n \normsq{\a_b} \nonumber \\
&=& \sum_{b=1}^n (a_{b,1} - a_{b,2})^2
\eneqn 

\noindent where the last equality follows from the fact found in Equation~\ref{eq:vector_fact}.

We now repeat this computation for the quadratic form for the B design. Here, we now redefine $\a_b := \v_b -  \onevec \v_b^\top \onevec / n_b$ and everything else remains the analogous so the previous steps 2, 3 are omitted.

\bneqn
\v^\top \bSigma_{BL(B)} \v  &=& \sum_{b=1}^B \v_b^\top \B_b \v_b \nonumber\\
&=& \sum_{b=1}^B \a_b^\top \parens{\frac{n_b}{n_b - 1}\I - \frac{1}{n_b - 1}\onevec\onevec^\top} \a_b \nonumber\\ \label{eq:n_b}
&=& \frac{n_b}{n_b - 1} \sum_{b=1}^B \normsq{\a_b}  \\
&=& \frac{1}{n_b - 1} \sum_{b=1}^B \sum_{1 \leq i < j \leq n_b} (a_{b,1} - a_{b,2})^2 \eqncomment{by Equation~\ref{eq:vector_fact}} \nonumber\\
\label{eq:B_mse}
&=& \frac{1}{n_b - 1} \sum_{b=1}^B \Big(  (a_{b,1} - a_{b,2})^2 + (a_{b,1} - a_{b,3})^2 + \ldots + (a_{b,1} - a_{b,n_b})^2 + \\
&& (a_{b,2} - a_{b,3})^2 + (a_{b,2} - a_{b,4})^2 + \ldots + (a_{b,2} - a_{b,n_b})^2 + \nonumber\\
&& + \ldots + (a_{b,n_b - 1} - a_{b,n_b})^2 \Big) \nonumber
%
\eneqn 


To compare the two expressions of Equations~\ref{eq:PM_mse} and \ref{eq:B_mse}, we can compare the quadratic form contribution of each block for the BL design (of size $n_b$) to the quadratic form contribution of the $n_b / 2$ blocks in the PM design. It is sufficient to demonstrate that the contribution from the PM design is bounded above by the contribution from the BL design for arbitrary block $b$. Without loss of generality, we consider the first block $b=1$ for the BL design (the first term in the sum of Equation~\ref{eq:B_mse}) which is

\bneqn\label{eq:block_contr_BL}
&& \frac{1}{n_b - 1} \big((a_{1,1} - a_{1,2})^2 + (a_{1,1} - a_{1,3})^2 + \ldots + (a_{1,1} - a_{1,n_b})^2 + \nonumber\\
&& (a_{1,2} - a_{1,3})^2 + (a_{1,2} - a_{1,4})^2 + \ldots + (a_{b,2} - a_{1,n_b})^2 + \nonumber\\
&& + \ldots + (a_{1,n_b - 1} - a_{1,n_b})^2 \big) \nonumber\\
&=& \frac{1}{n_b - 1} \big((v_{1} - v_{2})^2 + (v_{1} - v_{3})^2 + \ldots + (v_{1} - v_{n_b})^2 + \\
&& (v_{2} - v_{3})^2 + (v_{2} - v_{4})^2 + \ldots + (v_{2} - v_{n_b})^2 + \nonumber\\
&& + \ldots + (v_{n_b - 1} - v_{n_b})^2\big) \nonumber
\eneqn

\noindent where the equality follows from the fact that $\v$ has entries assumed ordered from smallest to largest and that we can decenter the $\a_1$ vector by adding its average. We now compare this to the contribution of the first $n_b$ subjects in the pairwise quadratic form expression which involves taking a partial sum of the expression in Equation~\ref{eq:PM_mse},

\bneqn\label{eq:block_contr_PM}
\sum_{b=1}^{n_b / 2} (a_{b,1} - a_{b,2})^2 &=& (a_{1,1} - a_{1,2})^2 + (a_{2,1} - a_{2,2})^2 + \ldots + (a_{n_b / 2,1} - a_{n_b / 2,2})^2 \nonumber \\ 
&=& (v_{1} - v_{2})^2 + (v_{3} - v_{4})^2 + \ldots + (v_{n_b - 1} - v_{n_b})^2
\eneqn

\noindent where the last equality again follows from the fact that $\v$ has entries assumed ordered from smallest to largest and that we can decenter the $\a_1$ vector by adding its average (which is a different average than for the BL expression). This sum resolves into pair differences only if $n_b / 2$ is an integer and thus this is the step that requires all $n_b$'s to be even.

We now demonstrate that the expression of Equation~\ref{eq:block_contr_PM} is less than or equal to the expression of Equation~\ref{eq:block_contr_BL} to complete the proof. Let $r := n_b - 1$. Equivalently, we wish to show that

\bneqn\label{eq:pyramid}
r(v_{1} - v_{2})^2 + r(v_{3} - v_{4})^2+ r(v_{5} - v_{6})^2  + \ldots + r(v_{n_b - 1} - v_{n_b})^2 &\leq& \nonumber\\
(v_{1} - v_{2})^2 + (v_{1} - v_{3})^2 + (v_{1} - v_{4})^2  + (v_{1} - v_{5})^2 + (v_{1} - v_{6})^2 + \ldots + (v_{1} - v_{n_b})^2 + && \\
 (v_{2} - v_{3})^2 + (v_{2} - v_{4})^2 + (v_{2} - v_{5})^2 + (v_{2} - v_{6})^2 + \ldots + (v_{2} - v_{n_b})^2 + &&\nonumber\\
 (v_{3} - v_{4})^2 + (v_{3} - v_{5})^2 + (v_{3} - v_{6})^2 + \ldots + (v_{3} - v_{n_b})^2 + &&\nonumber\\
 (v_{4} - v_{5})^2 + (v_{4} - v_{6})^2 + \ldots + (v_{4} - v_{n_b})^2 + &&\nonumber\\
(v_{5} - v_{6})^2 + \ldots + (v_{5} - v_{n_b})^2 + &&\nonumber\\
 \vdots && \nonumber\\
(v_{n_b - 1} - v_{n_b})^2 ~~&& \nonumber
\eneqn
\normalsize

\noindent We now proceed to demonstrate $r$ inequalities corresponding to each of the $r$ terms on the left hand side of the above. First, 

\bneqn\label{ineq:1}
r(v_{1} - v_{2})^2 \leq (v_{1} - v_{2})^2 + (v_{1} - v_{3})^2 + (v_{1} - v_{4})^2  + (v_{1} - v_{5})^2 + \ldots + (v_{1} - v_{n_b})^2
\eneqn
\normalsize

\noindent since the right hand side also has $r$ terms which are each less than or equal to the $(v_{1} - v_{2})^2$ due to the order of $\v$. Note that the right hand size corresponds to row one of the expression in Inequality~\ref{eq:pyramid}. Second,

\bneqn\label{ineq:2}
r(v_{3} - v_{4})^2 &\leq& \underbrace{\ingray{(v_{1} - v_{4})^2} + (v_{2} - v_{4})^2} + \\ 
&& (v_{3} - v_{4})^2  + (v_{3} - v_{5})^2   + (v_{3} - v_{6})^2+ \ldots + (v_{3} - v_{n_b})^2 \nonumber
\eneqn
\normalsize

\noindent for the same reasons given for why Inequality~\ref{ineq:1} is true. Note that the underbraced terms correspond to column three and the non-underbraced terms correspond to row three of the expression in Inequality~\ref{eq:pyramid}. The term in gray is double-counted between Inequalities~\ref{ineq:1} and \ref{ineq:2}. We will return to this point later. Third,

\bneqn\label{ineq:3}
r(v_{5} - v_{6})^2 &\leq& \underbrace{\ingray{(v_{1} - v_{6})^2} + (v_{2} - v_{6})^2 + \ingray{(v_{3} - v_{6})^2}  + (v_{4} - v_{6})^2} +   \\
&& (v_{5} - v_{6})^2 + \ldots + (v_{3} - v_{n_b})^2 \nonumber
\eneqn
\normalsize

\noindent for the same reason given for why Inequalities~\ref{ineq:1} and \ref{ineq:2} are true. Note that the underbraced terms correspond to column five and the non-underbraced terms correspond to row five of the expression in Inequality~\ref{eq:pyramid}. The first term in gray is double-counted between Inequalities~\ref{ineq:1} and \ref{ineq:3} and the second term in gray is double-counted between Inequalities~\ref{ineq:2} and \ref{ineq:3}. We imagine continuing this pattern of row/column additions through the $r-1$st term. The $r$th inequality then is

\bneqn\label{ineq:r}
r(v_{n_b - 1} - v_{n_b})^2 &\leq& \ingray{(v_{1} - v_{n_b})^2} + (v_{2} - v_{n_b})^2 + \ingray{(v_{3} - v_{n_b})^2} + (v_{4} - v_{n_b})^2 + \ingray{(v_{5} - v_{n_b})^2}   + \nonumber \\
&& \ldots + (v_{n_b - 1} - v_{n_b})^2
\eneqn
\normalsize

\noindent for the same reason given for why Inequalities~\ref{ineq:1}, \ref{ineq:2} and \ref{ineq:3} are true. Note that the terms on the right hand size correspond to the the $r$th column in Inequality~\ref{eq:pyramid}. The first term in gray is double-counted between Inequalities~\ref{ineq:1} and \ref{ineq:r}, the second term in gray is double-counted between Inequalities~\ref{ineq:2} and \ref{ineq:r}, the third term in gray is double-counted between Inequalities~\ref{ineq:3} and \ref{ineq:r}. There will be one double-counting every other term ending at term number $r-1$.

Note that throughout the $r$ inequalities, the terms in gray appear only once (hence they are double-counted and not triply-counted or more). We justify these double-countings given the following fact concerning three positive real numbers:

\bneqn\label{ineq:fact}
(a - b)^2 + (b - c)^2 \leq (a-c)^2~~ \text{for}~ 0 < a < b < c
\eneqn

\noindent For instance the first instance of double counting is for term $(v_1 - v_4)^2$ in Inequalities~\ref{ineq:1} and \ref{ineq:2}. Using the fact in Inequality~\ref{ineq:fact}, we know that

\beqn
(v_1 - v_2)^2 + (v_2 - v_4)^2 \leq (v_1 - v_4)^2
\eeqn


We now replace the  $(v_1 - v_4)^2$ term on the right hand side in Inequality~\ref{ineq:1} with $(v_1 - v_2)^2$, the first term on the left hand side above and replacing the $(v_1 - v_4)^2$ term on the right hand side in Inequality~\ref{ineq:2} with $(v_2 - v_4)^2$, the second term in the left hand side above. This effectively \qu{splits} the double-counted terms and does not break either inequality. Doing this splitting amongst all double-counted terms completes the proof. ~~$\blacksquare$

\subsection{PM is the Minimax Design}\label{app:pm_minimax}

Let $f(\v) := \v^\top \bSigmaw \v$. The function $f(\v)$ is convex because $\bSigmaw$ is positive semi-definite. Convexity implies that for $\v_1, \v_2 \in \mathcal{V}$, then $\max\braces{f(\v_1), f(\v_2)} \geq f(\alpha\v_1 + (1 - \alpha)\v_2)$ for all $\alpha \in \zeroonecl$. Thus, $\max_{{\bf v} \in {\cal V}} f({\bf v})$ is attained at corners of ${\cal V}$, where a point ${\bf c} \in {\cal V}$ is a corner if there exist no different ${\bf v}_1, {\bf v}_2 \in {\cal V}$ and $\alpha \in (0,1)$ such that ${\bf c}=\alpha {\bf v}_1 + (1-\alpha) {\bf v}_2$. The set of corners of ${\cal V}$ is ${\cal C}:= \{ (2,2,\ldots,2), (0,2,\ldots,2), \ldots, (0,0,\ldots,2), (0,0,\ldots,0)\}$. It follows that $\max_{{\bf v} \in {\cal V}} f({\bf v})= \max_{{\bf c} \in {\cal C}} f({\bf c})$. Since the point $(0,0,\ldots,2) \in {\cal C}$ and the diagonal elements of $\bSigmaw$ are all 1, we have $\max_{{\bf c} \in {\cal C}} f({\bf c}) \ge 2$.

For the PM design, $\bSigma_{PM} = \bSigma_{BL(n)}$ attains this lower bound for $f(\v)$, because for every ${{\bf c} \in {\cal C}}$, ${\bf c}^T \bSigma_{BL(n)} {\bf c} \in \braces{0, 2}$, depending on whether the number of 2's in ${\bf c}$ is even or odd. Therefore, $\max_{{\bf c} \in {\cal C}} {\bf c}^T \bSigma_{BL(n)} {\bf c}=2$ and thus

\beqn
\max_{\v \in \mathcal{V}} \braces{\v^\top \bSigma_n \v} = \min_{\W \in \mathcal{W}} \max_{\v \in \mathcal{V}} \braces{\v^\top \bSigmaw \v}
\eeqn

\noindent implying that PM provides the minimal MSE for worst case $\v$ as the quadratic form is the only design-dependent term in the MSE expression of Equation~\ref{eq:mse} of the main text. ~~$\blacksquare$

\subsection{Random Matching PM Performs the Same as BCRD}\label{app:pm_robust}

Under BCRD, 

\beqn
\prob{W_i = 1, W_j = 1} = n/(2n) \times (n - 1) / (2n - 1) = \half (n - 1) / (2n - 1)
\eeqn 

\noindent and under a random assignment in PM, 

\beqn
\prob{W_i = 1, W_j = 1} &=& \prob{W_i = 1} \cprob{W_j = 1, \angbrace{i,j} \in \mathcal{M}}{W_i = 1} + \\
&& \prob{W_i = 1} \cprob{W_j = 1, \angbrace{i,j} \notin \mathcal{M}}{W_i = 1} \\
&=& 0 + \\
&& \half (2n-2)/(2n-1)/2 = \half (n - 1) / (2n - 1).
\eeqn

\noindent Similarly, $\prob{W_i = k, W_j = \ell}$, for $k,\ell \in \{1,-1\}$, is equal in both PM (with random matching) and BCRD for all $i, j$. It follows, that the $\bSigma$ matrices are the same for PM and BCRD which implies their MSE's are identical for all $\v$ and all sample sizes. ~~$\blacksquare$

\subsection{Sufficient and necessary condition for PM to do worse than BCRD}\label{app:pm_not_robust}
	By Equation~\ref{eq:mse},
	\[
	\mse{\hat{\tau}_{BCRD}}-\mse{\hat{\tau}_{PM}}=\frac{1}{4n^2} \parens{ \v^\top \bSigma_{BCRD} \v - \v^\top \bSigma_{PM} \v}, 
	\] 
	and Equations~\ref{eq:vector_fact} and \ref{eq:PM_mse} imply that
	\beqn
	\v^\top \bSigma_{BCRD} \v - \v^\top \bSigma_{PM} \v=  \frac{1}{2n-1} \sum_{i<j} (v_i - v_j)^2 - \sum_{k=1}^n (v_{i_k}- v_{j_k})^2. 
	\eeqn

\noindent The condition of Equation~\ref{eq:cond} follows.  ~~$\blacksquare$

\subsection{Equality of Designs if the Covariates are Uninformative}\label{app:all_designs_equal}

Uninformative covariates imply the probabilities of treatment and control are equal for all subjects, hence $\v = c\onevec$ where $c \in (0,2)$. 

The only term that a design $\W$ can effect in the MSE of the estimator for $\betaThat$ (Equation~\ref{eq:mse}) is the quadratic form term,

\bneqn\label{eq:constant_quadratic_term}
\v^\top \bSigma_{\W} \v = c^2 \onevec^\top \bSigma_{\W} \onevec
\eneqn

\noindent The variance can be computed as

\beqn
\bSigma_{\W} = \expe{\W \W^\top} - \expe{\W} \expe{\W^\top} = \expe{\W \W^\top}
\eeqn

\noindent where the last equality follows by (A2). We then compute $\bSigma_{\W} \onevec$ from Equation~\ref{eq:constant_quadratic_term},

\beqn
\bSigma_{\W} \onevec = \parens{\oneover{\abss{\mathcal{W}}} \sum_{\ell=1}^{\abss{\mathcal{W}}} \w\w^\top} \onevec = \oneover{\abss{\mathcal{W}}} \sum_{\ell=1}^{\abss{\mathcal{W}}} \w (\w^\top\onevec) 
\eeqn

\noindent where $\abss{\mathcal{W}}$ denotes the number of vectors in the support of design $\W$. As $\w^\top\onevec = 0$ by (A1), the assumption that there are equal treatment and control subjects in all allocation in any design, the quadratic form term in the MSE is zero.~~$\blacksquare$
\end{document}